\documentclass[10pt]{amsart}
\usepackage{amsfonts,amsmath,amsthm,eucal}

\theoremstyle{plain}
\newtheorem{theorem}{Theorem}

\newtheorem{conj}{Conjecture}
\newtheorem{lemma}{Lemma}

\theoremstyle{definition}

\newtheorem{defn}{Definition}

\begin{document}
\title{R{\'e}nyi--Wehrl entropies as
  measures of localization in phase space}
\author[S. Gnutzmann]{Sven Gnutzmann}
\address{
  Sven Gnutzmann\\
  Department of Physics of Complex Systems\\
  The Weizmann Institute of Science\\
  Rehovot, Israel}
\email{sven@gnutzmann.net}
\author[K. \.Zyczkowski]{Karol \.Zyczkowski
}
\address{
  Karol \.Zyczkowski\\
  Centrum Fizyki Teoretycznej\\
  Polska Akademia Nauk\\
  Warszawa, Polska\\ {\it on leave from}
  Instytut Fizyki\\ Uniwersytet Jagiello{\'n}ski\\ Krak{\'o}w, Poland}
\email{karol@theta1.cft.edu.pl}
\begin{abstract}
  We generalize the concept of the Wehrl entropy of quantum states
  which gives a basis--independent
  measure of their localization in phase space.
  We discuss the minimal values and the typical values of these
  R{\'e}nyi--Wehrl entropies for pure states for spin systems.
  According to Lieb's conjecture the minimal values are provided
  by the spin coherent states. Though Lieb's conjecture remains
  unproven, we give new proofs of partial results that may be
  generalized for other systems.
  We also investigate random pure states and
  calculate the mean R{\'e}nyi--Wehrl entropies averaged over the
  natural measure in the space of pure quantum states.
\end{abstract}
\maketitle

\section{Introduction}

Analysis of localization properties of eigenstates of
a (chaotic) quantum mechanical system has lead to a
lot of insights on the behavior of classically chaotic
as well as disordered quantum systems.
The localization properties are usually characterized
by their eigenvector distribution
\cite{kus:88,bohigas:91,izrailev:90,haake:90}, the entropic
localization length \cite{casati:90} or the
 (inverse) participation ratio
\cite{heller:87}. All these quantities, however, are based
on the expansion of an eigenstate in some given
orthonormal eigenbasis, which mostly can be chosen arbitrarily.
If one chooses (with some bad will) the eigenbasis of the
system, all these quantities carry no information whatsoever.

Less known but very useful are measures of
localization that do not depend on some arbitrary choice.
These can be defined using
generalized coherent states \cite{perelomov:86},
which  provide a connection of
the quantum dynamics to the classical phase space.
They have been defined canonically for several examples of
classical phase spaces using the algebraic construction of Perelomov
and Gilmore
\cite{perelomov:86,gilmore:72} and provide a representation of
a quantum mechanical state by a positive normalized and bounded function
on phase space -- the so called Husimi function \cite{husimi:40}.
Let us mention in passing that the Husimi function was
successfully applied to study dynamical properties of quantized
chaotic systems \cite{takahashi:86,zyczkowski:87}.

The Wehrl entropy has been defined as measure of localization of the
Husimi function \cite{wehrl:78,wehrl:91}. This has the advantage of being
independent of the choice of some orthonormal basis \cite{kus:91}.
Probably even more
important is the fact that localization properties
of quantum mechanical states are measured in the
corresponding classical phase space
and can be connected to the classical dynamics
\cite{weinmann:99,varga:99,zyczkowski:2001}.
Generalized measures of localization are the moments of the Husimi
function or R{\'e}nyi--Wehrl entropies. The second moment has recently
ben proposed as a definition of complexity of quantum states \cite{sugita:2001}

The most localized quantum mechanical states are
those which minimize the Wehrl entropy.
For Glauber coherent states of
a harmonic oscillator \cite{glauber:63a,glauber:63b,sudarshan:63}
Wehrl conjectured
and Lieb proved that the state minimizing the entropy is
again a coherent state \cite{lieb:78}. Lieb also conjectured that
a similar statement should be true for the angular
momentum (spin) coherent states \cite{lieb:78}. This conjecture
remains unproven so far. There
are two ways to generalize the conjectures of Wehrl
and Lieb. The first is to generalize it to all systems
which allow for Gilmore--Perelomov coherent states.
Secondly one may generalize the measure of
localization itself (generalized R{\'e}nyi--Wehrl entropies).
In this work we discuss the latter
generalization and show that
the accordingly generalized Lieb--Wehrl conjecture
for the
coherent states can be proved for some of the generalized entropies.

This paper is organized as follows. After a
very brief introduction to Perelomov coherent states
we define in section \ref{sec:renyi-wehrl}
the R{\'e}nyi--Wehrl entropies
and  present the generalized Lieb--Wehrl conjectures.
In section 3 we compute the generalized Wehrl entropy
for the $SU(2)$ coherent states and prove that
they correspond to local minima of these entropies.
Moreover, we define the entropy reducing maps and using this notion
we provide a new proof of the generalized Lieb conjecture for
the R{\'e}nyi--Wehrl entropies of an integral order. The action of
rotations
on the Bargmann function, used in the proof, is described in Appendix A.
In section 4 we compute the mean R{\'e}nyi--Wehrl entropy
obtained by averaging over the natural measure on the space of pure
states. The paper is concluded in section 5.

\section{Generalized R{\'e}nyi--Wehrl entropies}
\label{sec:renyi-wehrl}

\subsection{Gilmore--Perelomov coherent states}
\label{subsec:perelomov}

The construction of Gilmore--Pere\-lomov
coherent states has  been described in
\cite{gilmore:72,perelomov:86}. Spin coherent states,
related to the rotation group $SU(2)$,
have been introduced in
\cite{radcliffe:71,arecchi:72}. Coherent states for other
compact groups ($SU(3)$ and $SU(n)$)
have been recently discussed
\cite{gitman:93,gnutzmann:98,nemoto:2000}.
Here we only give some of the general properties.  In
section \ref{subsec:spincoherent} we provide a more detailed
account on spin coherent states.

Let ${\mathcal M}$ be the classical phase space of some physical
system and $V$ the Hilbert space of the quantized
system. The coherent states of such a system
form a set of normalized states
$\left\{|\gamma\rangle\right\}_{\gamma\in {\mathcal M }}\subset V$
with the following properties:
\begin{itemize}
  \item[] {\bf CS.I}\\
    There is a one-to-one mapping
    ${\mathcal M} \longrightarrow V:\gamma \mapsto |\gamma\rangle$,
    such that a unique vector in Hilbert space
    corresponds to each point in the classical phase space.
  \item[] {\bf CS.II}\\
    For some set $\left\{O_i\right\}$ of observables the
    expectation values $\langle \gamma|O_i|\gamma\rangle $
    match the values of the corresponding classical observable
    at the point $\gamma$ in phase space. This condition can only
    be fulfilled if the observables $O_i$ build a small set
    of well chosen operators. This set should be large enough
    to determine uniquely  the point $\gamma$ in phase space
    from the expectation values.
  \item[] {\bf CS.III}\\
    The coherent states minimize the Heisenberg uncertainty
    among all norma\-lized states. By this property coherent states are
    in a sense the most classical quantum states.
  \item[] {\bf CS.IV}\\
    There exist a resolution of unity in coherent states,
    i.e. the set of coherent states is (over)complete.
\end{itemize}

In the first three conditions the coherent states are defined
as a subset of Hilbert space such that this subset
is equivalent to the classical phase space in a physically
(and mathematically) reasonable way. The condition CS.IV
needs additional explanation. Being a continuum of
normalized states the coherent states cannot form
an orthogonal basis -- in general the scalar product
of two different coherent states does not vanish.
 Still they satisfy a
completeness relation in the form of a resolution of unity
\begin{equation}
  {\mathbb I} = c \int_{\mathcal M} d\mu(\gamma) |\gamma\rangle\langle \gamma|,
  \label{eq:completeness}
\end{equation}
where ${\mathbb I}$ is the identity operator, $d\mu(\gamma)$
denotes a volume
element in the phase space (canonically defined by its symplectic
structure) and $c$ stands for a normalization constant. This looks
much like
the completeness relation for an orthonormal basis, in which
the sum has been replaced by an integral over the phase space.
For this reason the set of coherent states is often
said to be overcomplete.

For a large class of classical phase spaces the set of coherent states
can be constructed following the group theoretic approach of
Perelomov and Gilmore \cite{perelomov:86,gilmore:72}.
 Its main ingredient
is a Lie group $G$ and an irreducible representation of that
Lie group in some Hilbert space $V$.
This construction is known so far for certain classes
of Lie groups, including the physically most relevant groups
\cite{perelomov:86}.
 The coherent states are
obtained by action of the Lie group $G$
on an initial state $|0\rangle$
chosen according to condition CS.III,
\begin{equation}
  |\gamma\rangle= g(\gamma)|0\rangle
  \label{eq:grouponcoherentstate}
\end{equation}
where $g(\gamma)\in G$. It can be shown that the classical phase
space is equivalent to $G/H$,  where $H$ is the subgroup of $G$
that leaves $|0\rangle$ invariant up to some phase,
(\;$h|0\rangle=e^{i\phi}|0\rangle$
for all $h\in H$). The relevant set of observables (condition CS.II)
is given by the generators of the group. In the case of a spin
the Lie group is $G=SU(2)$, whose generators satisfy the
commutation relations of the components of an angular
momentum operator.

One straightforward way to describe quantum mechanics in phase space
is to use the overlap of a state with a coherent state, $\psi(\gamma)=
\langle\gamma|\psi\rangle$. The square of the absolute value of
this function
is called the Husimi function \cite{husimi:40} (or $Q$-function)\footnote{
In the mathematical literature the Husimi function
is also called covariant
or Wick symbol.}.
\begin{equation}
  {\mathcal H}_{|\psi\rangle}(\gamma)= |\langle \gamma|\psi\rangle|^2.
  \label{eq:husimidef}
\end{equation}
One can easily show that the Husimi function is
non-negative, bounded\footnote{Actually there
are many ways to define functions on phase space that represent
a quantum state completely. The Husimi function is the only
one which is non-negative and bounded.}
\begin{equation}
  0\le {\mathcal H}_{|\psi\rangle}(\gamma)\le 1 ,
  \label{eq:husimibound}
\end{equation}
and normalized
\begin{equation}
  c\int_{\mathcal M} d\mu(\gamma) {\mathcal H}_{|\psi\rangle}(\gamma)=
  \langle\psi|\psi\rangle=1  .
  \label{eq:husiminorm}
\end{equation}
One is tempted to think of the Husimi function as a
probability density on the phase space. However the rules
for calculating expectation values of some observable
using the Husimi function are non-classical. Still the
label quasiprobability has been introduced in this context
and in a properly defined classical
limit $c {\mathcal H}_{|\psi\rangle}$
becomes a usual classical probability distribution.

Due to the overcompleteness of the coherent states one can actually show
that the Husimi function determines the state $|\psi\rangle$ up to
some phase. This contrasts the properties of the expansion
of a state in an orthonormal basis, in which the phase
information is lost by squaring the expansion coefficients.
This is actually
the reason,  why expansions in coherent states are much
more powerful than
expansions in some orthonormal basis
for defining useful measures of localization.

In the following we will restrict ourselves to pure states.
Generalizations to mixed states are straightforward.
For the statistical operator
$\rho=\sum p_i |\psi_i\rangle\langle \psi_i |$
the Husimi function is
${\mathcal H}_{\rho}(\gamma)=\langle \gamma| \rho|\gamma\rangle$.
Again it is possible to reconstruct the statistical operator
$\rho$ from its Husimi function ${\mathcal H}_{\rho}(\gamma)$.

\subsection{Definition of the generalized R{\'e}nyi--Wehrl entropies}
\label{subsec:entropydef}

Wehrl has defined the phase space entropy of a quantum state
$|\psi\rangle$ as the continuous (Boltzmann) entropy
of its Husimi function \cite{wehrl:78},
\begin{equation}
  S_{|\psi\rangle}=- c\int_{\mathcal M} d\mu(\gamma)
   {\mathcal H}_{|\psi\rangle}
  \log {\mathcal H}_{|\psi\rangle}.
  \label{eq:wehrldef}
\end{equation}
The generalized Wehrl entropies carry the R{\'e}nyi index $q > 0$ and
are defined as
\begin{equation}
  S^{(q)}_{|\psi\rangle}=\frac{1}{1-q}\log W^{(q)}_{|\psi\rangle} ,
  \label{eq:renyiwehrldef}
\end{equation}
with the 'moment' functions
\begin{equation}
  W^{(q)}_{|\psi\rangle}=c
  \int_{\mathcal M} d\mu(\gamma)
  \left({\mathcal H}_{|\psi\rangle}\right)^q .
  \label{eq:momdef}
\end{equation}
In the limit $q\rightarrow 1$ one gets back the original Wehrl
entropy,
\begin{equation}
  \lim_{q \to 1}  S_{|\psi\rangle}^{(q)}=S_{|\psi\rangle},
  \label{eq:wehrllimit}
\end{equation}
so for consistency the Wehrl entropy
$S_{|\psi\rangle}$ will be denoted by
$S_{|\psi\rangle}^{(1)}$.

The R{\'e}nyi--Wehrl entropy is a non-increasing function of the parameter
$q$
\begin{equation}
  q_2 > q_1 \qquad \Longrightarrow \qquad S^{(q_2)}_{|\psi\rangle} \le
  S^{(q_1)}_{|\psi\rangle}.
  \label{eq:monotonicity}
\end{equation}
For compact phase spaces one can put $\int_{\mathcal M}
d\mu(\gamma)=1$. Then $c=N$ is the finite dimension of Hilbert
space. The function
\begin{equation}
  {\mathcal I}_{|\psi\rangle}(\gamma)=1-{\mathcal H}_{|\psi\rangle}(\gamma)
  \label{eq:Idef}
\end{equation}
is also bounded $0 \le {\mathcal I}_{|\psi\rangle}(\gamma) \le 1$
and normalizable. It thus makes sense to define
the following dual Wehrl entropy
\begin{equation}
  Z_{|\psi\rangle}=-\frac{N}{N-1}\int_{\mathcal M}
  d\mu(\gamma) {\mathcal I}_{|\psi\rangle}
  \log {\mathcal I}_{|\psi\rangle}
  \label{eq:dualwehrldef}
\end{equation}
and its R{\'e}nyi--Wehrl generalization
\begin{equation}
  Z^{(q)}_{|\psi\rangle}=\frac{1}{1-q} \log Y^{(q)}_{|\psi\rangle}
  \label{eq:dualrenyiwehrldef}
\end{equation}
with the dual moments
\begin{equation}
  Y^{(q)}_{|\psi\rangle}=\frac{N}{N-1}\int_{\mathcal M} d\mu(\gamma)
  \left({\mathcal I}_{|\psi\rangle}\right)^q.
  \label{eq:dualmomdef}
\end{equation}
Making use of the monotonicity of the logarithm
and the definition of the entropies it is easy to show that
for any two pure states $|\psi_1\rangle$ and $|\psi_2\rangle$
and any $q>1$ one has
\begin{equation}
  W^{(q)}_{|\psi_1\rangle}>W^{(q)}_{|\psi_2\rangle}
  \;\;\left(Y^{(q)}_{|\psi_1\rangle}>Y^{(q)}_{|\psi_2\rangle}\right)\qquad
  \Leftrightarrow \qquad
  S^{(q)}_{|\psi_1\rangle}<S^{(q)}_{|\psi_2\rangle}
  \;\;\left(Z^{(q)}_{|\psi_1\rangle}<Z^{(q)}_{|\psi_2\rangle}\right)
  \label{eq:inequality}
\end{equation}
while for $q<1$ one has $ W^{(q)}_{|\psi_1\rangle}>W^{(q)}_{|\psi_2\rangle}
\Leftrightarrow S^{(q)}_{|\psi_1\rangle}>S^{(q)}_{|\psi_2\rangle}$.

All the entropies $S^{(q)}$ and $Z^{(q)}$ provide measures
of localization of a quantum state in phase space.
Since the participation ratio is a very useful entropy-like
characteristic  of uniformity of a probability distribution,
we also define its analogue in phase space:
the Wehrl participation ratio,
\begin{equation}
  R_{|\psi\rangle}=
  \frac{1}{W^{(2)}_{|\psi\rangle}}=\exp S^{(2)}_{|\psi\rangle},
  \label{eq:PRdef}
\end{equation}
 and its dual
\begin{equation}
  T_{|\psi\rangle}=
  \frac{1}{Y^{(2)}_{|\psi\rangle}}=\exp Z^{(2)}_{|\psi\rangle}.
  \label{eq:dualPRdef}
\end{equation}

All these quantities are invariant under the action of the Lie group
$G$, e.g.
\begin{equation}
  S^{(q)}_{|\psi\rangle}= S^{(q)}_{g|\psi\rangle}
  \label{eq:groupinvariance}
\end{equation}
where $g\in G$. This ensures that all these entropies do
depend only on the shape of the Husimi function and not
on the position of a point in phase space, in
vicinity of which they are localized.

\subsection{The generalized Lieb-Wehrl conjecture}
\label{subsec:lieb-wehrl-conjecture}

For compact phase space the R{\'e}nyi--Wehrl entropies
are  bounded $S^{(q)}_{|\psi\rangle}\le \log N$
(with the dimension of Hilbert space $N$).
This upper bound corresponds
to a mixed state with statistical
 operator $\rho=\frac{1}{N}\,{\mathbb I}$.
The maximal value $S^{(q)}_{\rm max}$
for the R{\'e}nyi--Wehrl entropies
among pure states is strictly smaller.
However, neither the exact value  $S^{(q)}_{\rm max}$ nor the
states corresponding to it are known in general.

The R{\'e}nyi-Wehrl entropies are positive,
$S^{(q)}_{|\psi\rangle}>0$. The minimal value $S^{(q)}_{\rm min}$
can be expected for coherent states.
For $S_{|\psi\rangle}$ of a harmonic oscillator this was
conjectured by Wehrl \cite{wehrl:78}  and proved by Lieb
\cite{lieb:78}. Lieb extended the conjecture to
spin ($SU(2)$) coherent states,
which remains unproven up till now.
It is tempting to expect that the
following generalization is true.
\begin{conj}[Lieb--Wehrl, generalized]
  \label{conj:renyi-wehrl}
  The minimal value $S^{(q)}_{\rm min}$ of the
   R{\'e}nyi--Wehrl entropies  for $q>0$
  are obtained for coherent states
  \begin{equation}
    S^{(q)}_{\rm min}=S^{(q)}_{{\rm coh}}.
    \label{eq:minentropy}
  \end{equation}
  For compact phase spaces the same is true for the dual R{\'e}nyi--Wehrl
  entropies $Z^{(q)}_{\rm min}=Z^{(q)}_{{\rm coh}}$.
\end{conj}

\section{R{\'e}nyi--Wehrl entropies for spin coherent states}
\label{sec:spin}

\subsection{Phase space quantum mechanics of a spin and
  SU(2) coherent states}
\label{subsec:spincoherent}

As it is well known, an angular momentum operator may be described by
three operators $J_z,J_+=J_x+i J_y,J_-=J_x - i J_y$, fulfilling the
commutation relations $\left[J_+, J_- \right]=2 J_z$ and $\left[J_z,
  J_\pm\right]=\pm J_\pm$.  They are the generators of the Lie group
$SU(2)$.  The Hilbert space $V_j$ has dimension $N=2j+1$, where $j$ is
the spin quantum number defined by $ J^2= j(j+1) {\mathbb I}$. An
orthonormal basis of $V_j$ is given by the eigenstates of the $z$-th
component of the spin, $J_z |m\rangle = m |m\rangle$ with
$m=-j,-j+1,\dots,j-1,j$.  The classical phase space for a spin is the
Bloch sphere $S^2$. For convenience we choose stereographic
coordinates on the sphere. A point on the sphere is then given by one
complex number $\gamma \in {\mathbb C}$ and has polar coordinates
given by
\begin{equation}
  {\rm cos}\,\theta= \frac{1-|\gamma|^2}{1+|\gamma|^2}\qquad
   {\rm and} \qquad
  e^{i\phi}= \frac{\overline{\gamma}}{|\gamma|}.
  \label{eq:polarcoord}
\end{equation}
The 'north-pole' has the coordinate $\gamma=0$ and the 'south-pole'
can only be reached in the limit $|\gamma|\rightarrow\infty$.

The state $|j\rangle$ is localized at the north-pole of the sphere and
can be shown to minimize an uncertainty relation \cite{delbourgo:77}.
Coherent states $|\gamma\rangle$ are obtained by rotating the state
$|j\rangle$ by some rotation matrix ${\mathcal R}(\gamma) \in SU(2)$.
Equivalently one may define first a set of non-normalized coherent
states by
\begin{equation}
  |\gamma )=e^{\gamma J_-}|j\rangle.
  \label{eq:defcoherent}
\end{equation}
Their normalized counterparts are, (note the difference in brackets
used)
\begin{equation}
  |\gamma\rangle= \frac{1}{\sqrt{{\mathcal K}(\gamma)}}|\gamma)
  \qquad \text{with} \qquad
  {\mathcal K}(\gamma)=(\gamma|\gamma )= \left( 1+ |\gamma|^2 \right)^{2j}.
  \label{eq:normcoherentstates}
\end{equation}
Obviously the state $|j\rangle$ is a coherent state with $\gamma=0$.

The resolution of unity is given by
\begin{equation}
  {\mathbb I} = \left( 2j+1\right) \int_{S^2} d\mu(\gamma)
  |\gamma\rangle\langle \gamma|
  = \left(2j+1\right) \int_{S^2} d\mu(\gamma) \frac{1}{{\mathcal K}(\gamma)}
  |\gamma)(\gamma|
  \label{eq:resolunity}
\end{equation}
with the rotation-invariant area element (Haar measure)
\begin{equation}
  d\mu(\gamma)=\frac{1}{\pi \left( 1+|\gamma|^2\right)^2}
  \,d{\rm Re}(\gamma)\,d{\rm Im}(\gamma)
\end{equation}
on the Bloch sphere.

Any state $|\psi\rangle$ is uniquely determined by its overlap with
coherent states. Thus, one may define the so called {\sl Bargmann
  functions} on the sphere
\begin{equation}
  \gamma\mapsto\psi(\gamma)=\langle\psi|\gamma)=
  \langle\psi|e^{\gamma J_-}|j\rangle .
  \label{eq:bargmanndef}
\end{equation}
As $ J_-^k |j\rangle = 0$ for $k>2j$ the Bargmann function is a finite
polynomial in $\gamma$
\begin{equation}
  \psi(\gamma)=\sum_{k=0}^{2j} f_k \gamma^k
  \label{eq:bargmann}
\end{equation}
with some complex coefficients $f_k$. The scalar product of two states
can be written in terms of their Bargmann functions as
\begin{equation}
  \langle \psi_1|\psi_2\rangle = (2j+1)\int d\mu(\gamma)
  {\mathcal K}^{-1}(\gamma)
  \psi_1(\gamma) \overline{\psi_2(\gamma)}.
  \label{eq:scalprodbargamann}
\end{equation}
The normalization of a state then leads to the condition
$\sum_{k=0}^{2j} \binom{2j}{k}^{-1}|f_k|^2=1$.  As the monomials
$\gamma^k$ ($k=0,1,2,\dots$) build an orthogonal (but not normal)
basis one also has $|f_k|^2\le\binom{2j}{k}$.  If the state is a
normalized coherent state, $|\psi\rangle= {\mathcal
  K}(\gamma')^{-1/2}|\gamma')$, the corresponding Bargmann function
reads
\begin{equation}
  \psi(\gamma)=\langle\gamma'|\gamma)=
  {\mathcal K}(\gamma')^{-1/2}(\gamma'|\gamma)=
  \frac{(1+\overline{\gamma'}\gamma)^{2j}}{(1+|\gamma'|^2)^j} \ .
  \label{eq:coherentbargmann}
\end{equation}
The limit $|\gamma'|\rightarrow \infty$ corresponds to the Bargmann
function for a coherent state on the south pole.  While $(\gamma'|$
does not exist in this limit, its Bargmann function is well behaved
and gives ${\mathcal K}(\gamma')^{-1/2}(\gamma'|\gamma)
\longrightarrow \gamma^{2j} e^{-4ij\alpha}$, where the phase $\alpha$
is defined by $\gamma'=|\gamma'|e^{i\alpha}$. The Bargmann function
for a coherent state on the 'north pole' is simply constant,
$\psi(\gamma)= {\mathcal
  K}(\gamma')^{-1/2}(\gamma'|\gamma)|_{\gamma'=0}=1$.

Every state $|\psi\rangle$ is defined up to some phase factor by the
$2j$ zeroes $\gamma^{(0)}_k$ of its Bargmann function by the {\sl
  stellar representation} \cite{majorana:32,bacry:74,penrose:89}
\begin{equation}
  \psi(\gamma)={\mathcal N}^{-1}
  \prod_{k=1}^{2j-m}(\gamma-\gamma^{(0)}_k) ,
  \label{eq:stellar}
\end{equation}
where ${\mathcal N}$ is some normalization constant and $m$ is the
number of zeroes at the south pole.\footnote{The limit when one of the
  $\gamma^{(0)}_k$ approaches infinity is well defined and reduces the
  rank of the polynomial.}  A coherent state is characterized by one
$2j$-fold degenerate zero of the Bargmann function.

The Husimi function of a pure state is closely related with its
Bargmann function,
\begin{equation}
  {\mathcal H}_{\psi}(\gamma)=
  \frac{|\psi(\gamma)|^2}{{\mathcal K}(\gamma) }.
  \label{eq:husimi}
\end{equation}
Just as the Bargmann function the Husimi function of a pure state is
completely determined by its $2j$ zeroes.  Therefore, the zeroes of
the Husimi and the Bargamann functions coincide for any pure state.

From Schr\"odinger's equation $i\hbar \frac{d}{dt}|\psi(t)\rangle= H
|\psi(t)\rangle$ with the hermitian Hamiltonian $H$ one gets an
evolution equation for the Bargmann function,
$\psi(\gamma;t)=\psi(t)(\gamma)$,
\begin{equation}
  i\hbar \frac{d}{dt} \overline{\psi(\gamma;t)}=(2j+1) \int
  d\mu(\gamma') {\mathcal K}(\gamma')^{-1} H(\overline{\gamma},\gamma')
  \overline{\psi(\gamma';t)}
  \label{eq:dbargmanndt}
\end{equation}
with
\begin{equation}
  H(\overline{\gamma},\gamma')=( \gamma | H | \gamma' )=\sum_{k,k'=0}^{2j}
  H_{kk'}\overline{\gamma}^k {\gamma'}^{k'}.
  \label{eq:hamiltonian}
\end{equation}
The hermiticity of $H$ is reflected in
$\overline{H(\overline{\gamma},\gamma')}=
H(\overline{\gamma}',\gamma)$   and $H_{kk'}=\overline{H_{k'k}}$.  For
the time dependence of the Husimi function this gives
\begin{equation}
  i \hbar \frac{d}{dt}{\mathcal H}_{|\psi\rangle}(\gamma;t)=
  \int d\mu(\gamma') \frac{2j+1}{{\mathcal K}(\gamma){\mathcal K}(\gamma')}
  \left( \psi(\gamma)
    H(\overline{\gamma},\gamma')\overline{\psi(\gamma')}
    -\psi(\gamma')H(\overline{\gamma}',\gamma)\overline{\psi(\gamma)}
  \right).
  \label{eq:dhusimidt}
\end{equation}
The integral equations \eqref{eq:dbargmanndt} and \eqref{eq:dhusimidt}
can in general be transformed into equivalent differential equations.
For our purposes the integral form is better suited.

\subsection{R{\'e}nyi--Wehrl entropies for spins}

It is usually quite difficult to give a closed expression
for the integrals $W^{(q)}_{|\psi\rangle}$ and $Y^{(q)}_{|\psi\rangle}$
for a general state $|\psi\rangle$. For the Wehrl entropy $S_{|\psi\rangle}$
the stellar representation helps and an explicit expression
was given by Lee \cite{lee:88,schupp:99}.

For a coherent state the integrals are straightforward (using
${\mathcal H}={\mathcal K}^{-1}$ for the coherent state on the north
pole $\gamma=0$) and all measures of localization can be
expressed analytically
\begin{align}
  W^{(q)}_{{\rm coh}}=&\frac{2j+1}{2qj+1} \ ,
  &
  Y_{{\rm coh}}^{(q)}=&\frac{\Gamma(q+1)\Gamma(\frac{1}{2j}+2)}{
    \Gamma(q+1+\frac{1}{2j})}\ , \\
  S_{{\rm coh}}=&\frac{2j}{2j+1} \ ,
  &
  Z_{{\rm coh}}=&
  \Psi(\frac{4j+1}{2j})+\boldsymbol{\gamma}-1\ , \\
  R_{{\rm coh}}=&\frac{4j+1}{2j+1} \ ,
  &
  T_{{\rm coh}}=&\frac{4j+1}{4 j}.
  \label{eq:coherententropies}
\end{align}
Here $\Psi(x)$ denotes the Digamma function and $\boldsymbol{\gamma}
=0.5772\dots$
is the Euler constant. For the eigenstates $|m\rangle$ ($m=-j,\dots,j$)
of $J_z$ one obtains
\begin{align}
  W^{(q)}_{|m\rangle}=&
  \frac{2j+1}{2qj+1} \binom{2j}{j-m}^q \binom{2qj}{q(j-m)}^{-1} \ ,
  \\
  S_{|m\rangle}=& \frac{2j}{2j+1} - \log \binom{2j}{j-m} +2j\,
  \Psi(2j+1) +
  \\
  &
  -(j+m)\Psi(j+m+1) -(j-m)\Psi(j-m+1) \ ,
  \nonumber\\
  R_{|m\rangle}=&\frac{4j+1}{2j+1}
  \binom{4j}{2(j-m)}\binom{2j}{j-m}^{-2}.
  \label{eq:jzentropies}
\end{align}
For $m=-j$ or $m=j$ this reproduces the results for coherent states.
For $m=0$ the state is localized on the equator of the Bloch sphere.
For large $N=2j+1$ the Wehrl entropy of this state increases
like $S_{|m=0\rangle}\sim
\frac{1}{2}\log N+ const$. As it was shown in \cite{zyczkowski:2001}
this behaviour is typical for the eigenstates of an integrable map on
the sphere. However, for a typical (random) state the Wehrl entropy
grows as $S_{|\psi\rangle}\sim \log N -const$, with a positive constant.
 Such behaviour is found for the eigenstates
of quantized chaotic maps like the periodically kicked top.
Asymptotically, both results
differ by a factor of two. This reflects the fact that the eigenstates
of chaotic systems are delocalized over entire sphere, while
the eigenstates of integrable systems typically cover a band
along a (possibly deformed) circle on the sphere.

For integer values of the R{\'e}nyi parameter, $q=2,3,\dots$ ($q=1$ is
trivial and gives the norm),  it is possible to give
an explicit result for the moments
\begin{equation}
  W^{(q)}_{|\psi\rangle}=\sum_{m=0}^{2qj}
  \frac{2j+1}{2qj+1}\binom{2qj}{m}^{-1}
  \left|
    {\sum_{i_1,i_2,\dots,i_q}}
    \prod_{k=1}^q f_{i_k} \right|^2   \ ,
  \label{eq:explicitmoments}
\end{equation}
where the coefficients $f_k$ are given by \eqref{eq:bargmann}
and the inner sum goes from $0$ to $2j$ for each $i_k$ with
the restriction $\sum_{k=1}^q i_k=m$.

\subsection{Time dependence of the R{\'e}nyi--Wehrl entropies}
\label{subsec:timedependence}

If a time-dependent state
$|\psi(t)\rangle=e^{-i\frac{Ht}{\hbar}}|\psi(0)\rangle$
obeys the Schr\"odinger's equation, the R{\'e}nyi--Wehrl entropies
$S^{(q)}_{|\psi(t)\rangle}$
will generally vary in time. However, if the Hamiltonian is of the
form
$H= a J_z + b J_+ + \overline{b} J_-$
 ($a\in {\mathbb R}$, $c\in {\mathbb C}$),
the evolution of a state corresponds to a rotation
$e^{-i\frac{Ht}{\hbar}}={\mathcal R}\in SU(2)$,
the coherent states remain coherent and
 all R{\'e}nyi--Wehrl entropies
are constant in time. In general  $S^{(q)}_{|\psi(t)\rangle}$
is quasiperiodic. The rate of change of the entropies and moments
is given by
\begin{align}
  \frac{d}{dt} S_{|\psi(t)\rangle}=&
  -(2j+1)\int d\mu(\gamma)
  \log {\mathcal H}_{|\psi\rangle}(\gamma;t)
  \left[
    \frac{d}{dt}{\mathcal H}_{|\psi\rangle}(\gamma;t)
  \right] \ ,
  \label{eq:ddtS}
  \\
  \frac{d}{dt} S^{(q)}_{|\psi(t)\rangle}=&
  \frac{1}{(1-q) W^{(q)}_{|\psi(t)\rangle}}
  \left[
    \frac{d}{dt} W^{(q)}_{|\psi(t)\rangle}
  \right]   \ ,
  \label{eq:ddtSq}
  \\
  \frac{d}{dt} W^{(q)}_{|\psi(t)\rangle}=&
  q (2j+1) \int d\mu(\gamma)
  \left(
    {\mathcal H}_{|\psi\rangle}(\gamma;t)
  \right)^{q-1}
  \left[
    \frac{d}{dt}{\mathcal H}_{|\psi\rangle}(\gamma;t)
  \right]    \ ,
  \label{eq:ddtW}
  \\
  \frac{d}{dt} Z_{|\psi(t)\rangle}=&
  \frac{2j+1}{2j}\int d\mu(\gamma)
  \log(1- {\mathcal H}_{|\psi\rangle}(\gamma;t))
  \left[
    \frac{d}{dt}{\mathcal H}_{|\psi\rangle}(\gamma;t)
  \right]     \ ,
  \label{eq:ddtZ}
  \\
  \frac{d}{dt} Z^{(q)}_{|\psi(t)\rangle}=&
  \frac{1}{(1-q) Y^{(q)}_{|\psi(t)\rangle}}
  \left[
    \frac{d}{dt} Y^{(q)}_{|\psi(t)\rangle}
  \right]  \ ,
  \label{eq:ddtZq}
  \\
  \frac{d}{dt} Y^{(q)}_{|\psi(t)\rangle}=&
  - q \frac{2j+1}{2j}\int d\mu(\gamma)
  \left(1-
    {\mathcal H}_{|\psi\rangle}(\gamma;t)
  \right)^{q-1}
  \left[
    \frac{d}{dt}{\mathcal H}_{|\psi\rangle}(\gamma;t)
  \right]   \ .
  \label{eq:ddtY}
\end{align}
Pure states, for which 
these time derivatives vanish for
every Hamiltonian $H$ will be called {\sl extremal}.
It will be shown below that the coherent states belong to this class.
For an extremal state it is also interesting to calculate
the second derivatives which we only need for the moment functions
\begin{align}
  \frac{d^2}{dt^2} W^{(q)}_{|\psi(t)\rangle}=&
  q(q-1)(2j+1)
  \int d\mu(\gamma)\left(
    {\mathcal H}_{|\psi\rangle}(\gamma;t)\right)^{q-2}
  \left[
    \frac{d}{dt}{\mathcal H}_{|\psi\rangle}(\gamma;t)
  \right]^2
  \nonumber\\
  &+ q(2j+1)
  \int d\mu(\gamma)
  \left(
    {\mathcal H}_{|\psi\rangle}(\gamma;t)\right)^{q-1}
  \left[
    \frac{d^2}{dt^2}{\mathcal H}_{|\psi\rangle}(\gamma;t)
  \right]  \ ,
  \label{eq:d2dt2W}
  \\
  \frac{d^2}{dt^2} Y^{(q)}_{|\psi(t)\rangle}=&
  q(q-1)\frac{2j+1}{2j}
  \int d\mu(\gamma)\left(
    1-{\mathcal H}_{|\psi\rangle}(\gamma;t)\right)^{q-2}
  \left[
    \frac{d}{dt}{\mathcal H}_{|\psi\rangle}(\gamma;t)
  \right]^2
  \nonumber\\
  &- q(2j+1) \int d\mu(\gamma)
  \left(
    1-{\mathcal H}_{|\psi\rangle}(\gamma;t)\right)^{q-1}
  \left[
    \frac{d^2}{dt^2}{\mathcal H}_{|\psi\rangle}(\gamma;t)
  \right] .
  \label{eq:d2dt2Y}
\end{align}

\subsection{The generalized Lieb--Wehrl conjecture for spins}
\label{subsec:status}

The generalized Lieb--Wehrl conjecture stated in section
\ref{subsec:lieb-wehrl-conjecture} is not easy to
prove even in  the simplest case of the $SU(2)$ coherent states and the
Bloch
sphere.
 The original conjecture of Lieb concerning the Wehrl entropy
$S_{|\psi\rangle}$
(i.e. the case $q=1$)  has been proved only for
small spin quantum numbers $j=1$ and $j=3/2$
\cite{schupp:99,scutaru:79}\footnote{The case $j=1/2$ is
trivial as every pure state is a coherent state.}.
It is however possible to show that all states sufficiently near to
a coherent state have larger R{\'e}nyi--Wehrl entropies.
\begin{theorem}
  The coherent states minimize the entropies $S^{(q)}$ and $Z^{(q)}$
  locally for any $q > 0$.
  The moments $W^{(q)}$ and $Y^{(q)}$
  are locally maximized if $q>1$ and locally
  minimized if $0<q<1$.
\end{theorem}
\begin{proof}
  In the Hilbert space $V_j$ any two normalized pure states can
  be transformed into another by a unitary operator $U$,
  $|\psi_1\rangle=U|\psi_2\rangle$. Also, for any unitary operator
  on $V_j$ there is a hermitian operator $H$ such that
  $U=e^{-iH}$.
  In this proof we think of $H$ as the Hamiltonian of a dynamical
  quantum system and $|0\rangle$ as the initial condition. 
  Now for any state $|\psi\rangle$
  in a neighborhood of
  the coherent state on the north pole, $|0\rangle$, there
  is a Hamiltonian, such that after some time $t$ the
  dynamical state of the system is $|\psi\rangle$.

  It suffices to prove the statement on the moment functions.
  Due to invariance under rotation one may choose the coherent state
  on the north pole ($\gamma=0$) with the Husimi function
  ${\mathcal H}=\frac{1}{(1+|\gamma|^2)^{2j}}$. Choose this state
  as initial state for Schr\"odinger's equation with some
  arbitrary Hamiltonian $H$. Using equations \eqref{eq:ddtW} and
  \eqref{eq:ddtY} we conclude that at $t=0$ the first derivative of
  these moments vanish for every Hamiltonian $H$. In more tedious
  calculations for
  the second derivatives at $t=0$ one uses equations \eqref{eq:d2dt2W}
   and   \eqref{eq:d2dt2Y}.
  For $q>1$ ($0 < q < 1$) the second derivatives of the moments are
  either negative (positive) or they vanish. The Hamiltonians $H$ which
  lead to the vanishing second derivative either have the coherent
  state on the north pole as an eigenstate or are
  locally equivalent to a rotation. Thus in all cases with
  vanishing second derivative the state remains within the
  manifold of coherent states for infinitesimal times.
\end{proof}

To complete the proof of the Lieb--Wehrl conjecture
for the Bloch sphere one needs to show that $S^{(q)}_{|\psi\rangle} >
S^{(q)}_{\rm coh}$ for any non-coherent state $|\psi\rangle$.
\begin{defn}
  We call a general norm--preserving map
  ${\mathcal F}:V_j\rightarrow V_j$ and ${\mathcal F}:|\psi\rangle \mapsto
  {\mathcal F}(|\psi\rangle)$ with ${\mathcal F} (\alpha |\psi\rangle)= 
  \alpha{\mathcal F} (|\psi\rangle)$
  {\it entropy reducing} if we have
  $S^{(q)}_{|\psi\rangle} \ge S^{(q)}_{{\mathcal F}(|\psi\rangle)}$
  for every $|\psi\rangle \in V_j$.
\end{defn}
The map ${\mathcal F}$ need not be one-to-one or linear, so it need not have
any physical relevance. However, finding such an entropy reducing map
would be a step toward the proof of the Lieb conjecture.
Loosely  speaking we are looking for a sequence of contracting
transformations in the space of all pure states
(complex projective space ${\mathbb C}P^{N-1}$),
with an attractor at the set of coherent states,
(${\mathbb C}P^{1}\approx S^2 \subset {\mathbb C}P^{N-1}$).

We shall discuss the following candidates for such entropy reducing
maps, defined by their action on the Bargmann function

Let ${\mathcal F}_1:V_j\rightarrow V_j$
be the map defined in the stellar representation by
\begin{equation}
  \psi(\gamma)={\mathcal N}^{-1}\prod_k (\gamma-\gamma^{(0)}_k)
  \mapsto {\mathcal F}_1\circ \psi(\gamma)=
  {{\mathcal N}'}^{-1}\prod_k (\gamma-|\gamma^{(0)}_k|) .
  \label{eq:F1}
\end{equation}

The geometrical meaning
of this map is to rotate all zeroes independently around the z-Axis
until they meet the Greenwich meridian defined by $\phi=0$.
Rotating the zeroes independently is in general a non-unitary and a
noninvertible map.
The above map ${\mathcal F}_1$ may be supplemented by another
map ${\mathcal F}_1'$, which moves all zeros located
along the meridian, into the direction of a certain point belonging to
the meridian, say its center at $\theta=\pi/2$.
Then ${\mathcal F}_1'={\mathcal F}_1 \circ {\mathcal R}_x(\pi/2)$
where ${\mathcal R}_x(\pi/2)$ is a quarter rotation around the x-axis
and the composition ,
${\mathcal F}'_1 \circ {\mathcal F}_1={\mathcal F}_1
\circ {\mathcal R}_x(\pi/2)\circ {\mathcal F}_1$,
transforms all zeros of the Husimi function into one point, or in other
words, maps any pure state into a coherent state.

Another map ${\mathcal F}_2:V_j\rightarrow V_j$
can similarly be defined for
the coefficients
\begin{equation}
   \psi(\gamma)=\sum_{k=0}^{2j} f_k \gamma^k
   \mapsto {\mathcal F}_2\circ \psi(\gamma)=
   \sum_{k=0}^{2j} |f_k| \gamma^k.
   \label{eq:F2}
\end{equation}
There is no geometrical
interpretation like before. However, all zeroes of ${\mathcal
F}_2(|\psi\rangle)$
come in complex conjugate pairs and negative real part.

Both ${\mathcal F}_1$ and ${\mathcal F}_2$
are not invariant under unitary
rotions ${\mathcal R}\in SU(2)$ of the states
\begin{equation}
  {\mathcal F}_i \circ {\mathcal R} \neq {\mathcal R}\circ{\mathcal F}_i.
\end{equation}
Also, the map ${\mathcal F_i}\circ{\mathcal R}$ is entropy reducing if and only if
${\mathcal F}_i$ is. In general, any map ${\mathcal F}$ that increases
the density of zeroes of the Bargmann function in a certain region
of phase space is a candidate for an entropy reducing map.
This follows from the observation that a coherent state
has only one $2j$-fold degenerate zero.
As such an alternative one could consider another
transformation ${\mathcal F}_3$,
 pushing independently each zero
into the direction determined by the barycenter of all zeros,
${\vec{r}}_B:=\sum_{i=1}^N {\vec{r}}_i$, where
${\vec{r}}_i$ denotes the unit vector pointing
to the zero $\gamma_i$. In
the generic case the radius of the barycenter is nonzero, and its
direction distinguishes a certain point in the sphere.
This transformation reduces the simplified Monge distance
of any pure state to the set of coherent
states, which is defined by the minimal sum of
distances (measured as angles on the sphere)
of all $N$ zeros to any point on the sphere \cite{ZS00}.

Not being able to provide a general proof
that the maps ${\mathcal F}_i$ are entropy reducing, we
can give the following result.
\begin{lemma}
  \label{lemma1}
  For $q=2,3,\dots$ and $|\psi'\rangle={\mathcal F}_2(|\psi\rangle)$
  one has $S^{(q)}_{|\psi'\rangle} \le S^{(q)}_{|\psi\rangle}$ 
  or, equivalently,  $W^{(q)}_{|\psi'\rangle} \ge W^{(q)}_{|\psi\rangle}$.
  Equality holds if and only if the phases of the coefficients
  obey $f_k=e^{i\alpha k}$ for some fixed angle $\alpha$.
\end{lemma}
\begin{proof}
  It suffices to prove the statement on the moments $W^{(q)}$.
  In equation \eqref{eq:explicitmoments} the moments were
  calculated in the form
  $W^{(q)}_{|\psi\rangle}=\sum_{m=0}^{2qj}
  a_{qjm}
  |\sum_i z_i|^2$ with coefficients $a_{qjm}>0$.\\
  Now $W^{(q)}_{|\psi'\rangle} \ge W^{(q)}_{|\psi'\rangle}$ follows
  immediately from
  $|\sum_i z_i|^2\le (\sum_i |z_i|)^2$. The statement on
  equality follows from the condition $|\sum_i z_i|^2 = (\sum_i |z_i|)^2$.
\end{proof}
Equipped with this lemma we may now state a proof of
the generalized Lieb--Wehrl conjecture for the integer values of the R{\'e}nyi
parameter \footnote{An alternative
proof of this fact has been already provided by Schupp
\cite{schupp:99}.}.
\begin{theorem}
  For $q=2,3,\dots$ the inequalities $S^{(q)}_{\rm coh}\le
  S^{(q)}_{|\psi\rangle}$ and $W^{(q)}_{\rm coh}\ge
  W^{(q)}_{|\psi\rangle}$ hold for every state $|\psi\rangle \in V_j$
  with strict inequality for non-coherent states.\\
  Similarly $Z_{\rm coh}\le Z_{|\psi\rangle}$.
\end{theorem}
\begin{proof}
  The statements on $S^{(q)}$ and $W^{(q)}$ are equivalent.
  The statement on $Z_{|\psi\rangle}$ can
  be reduced to the statements on $W^{(q)}$
  using
  \begin{equation}
    Z_{|\psi\rangle}=1-\frac{1}{2j}\sum_{n=2}^{\infty}
      \frac{W^{(n)}_{|\psi \rangle}}{n(n-1)}.
  \end{equation}

  For any non-coherent state $|\psi\rangle$ described
  by its Bargmann function $\psi(\gamma)$ we will construct
  a state $\psi'(\gamma)$ such that
  $W^{(q)}_{|\psi\rangle}<W^{(q)}_{|\psi'\rangle}$, using
  the entropy reducing map ${\mathcal F}_2$
  defined in \eqref{eq:F2} and rotations ${\mathcal R} \in SU(2)$.
  As $W^{(q)}_{|\psi\rangle}$ is a bounded function on a compact
  manifold (the projective space ${\mathbb P}V_j$) and the coherent
  states provide a local maximum, finding such a state
  is equivalent to showing that the coherent states are a global
  maximum and that every global maximum is at a coherent state.

  In general we have
  $W^{(q)}_{|\psi\rangle} \le W^{(q)}_{{\mathcal F}_2(|\psi\rangle)}$ with
  strict inequality if ${\mathcal F}_2(|\psi\rangle)$ is a coherent
  state while $|\psi\rangle$ is not. So there are no
  noncoherent states that are mapped to a coherent state
  by ${\mathcal F}_2$ without increasing $W^{(q)}$.
  It follows that we may assume in the following
  that all coefficients $f_k$ of
  $\psi(\gamma)=\sum_{k=0}^{2j} f_k \gamma^k$ are real
  and nonnegative, $f_k \ge 0$ (and we still assume that $\psi(\gamma)$
  is not a coherent state).
  Since $\psi(\gamma)$ does not describe a coherent state,
  there is no $z\in {\mathbb R}$ such that
  $f_k=\binom{2j}{k} \frac{z^k}{(1+|z|^2)^{2j}}$.
  Further action of ${\mathcal F}_2$ cannot increase
  the value of $W^{(q)}$.

  We now show that the combined action of
  a suitable rotation
  ${\mathcal R}(\alpha,\phi) \in SU(2)$ as given by
  \eqref{eq:rotmat} and \eqref{eq:representrot}
  followed by ${\mathcal F}_2$  gives  the desired result.
  The action of rotations on Bargmann functions is described in the
  appendix \ref{app:rotations}.
  It suffices to consider the transformation
  ${\mathcal R}_x:={\mathcal R}(\alpha=x,\phi=0)$
  with real $x$, which represents the rotation around
  the y-axis.
  The rotated Bargmann function
  for an arbitrary $x$ is then given by
  \begin{align}
    {\mathcal R}_x\circ \psi(\gamma)=&
    \frac{(1+x\gamma)^{2j}}{(1+x^2)^j}
    \psi \left(\frac{\gamma+x}{1+x\gamma}\right)
    \\
    =&
    \frac{1}{(1+x^2)^j}\sum_{k=0}^{2j} f_k (\gamma+x)^k(1+x\gamma)^{2j-k}
    \nonumber
    \\
    =&\sum_{k=0}^{2j} \tilde{f}_k[x] \gamma^k.
    \nonumber
  \end{align}
  Such a rotation leaves the moments $W^{(q)}$ invariant. We now
  want to find a value $x_0$ so that the combined action 
  ${\mathcal F}_2{\mathcal R}_{x_0}$ increases these moments
  by a finite amount when applied to $\psi(\gamma)$.
  
  We consider the coefficient
  \begin{equation}
    \tilde{f}_0[x]=\frac{1}{(1+x^2)^j}\sum_{k=0}^{2j} f_k x^k
  \end{equation}
  of the rotated Bargmann function
  as a function of $x$. Since all the coefficients
  $f_k$ are positive and $\tilde{f}_0(x)\rightarrow 0$
  when $x\rightarrow \infty$, $\tilde{f}_0[x]$ has a maximum value for
  some $x_{\rm max}\ge 0$. As $\psi(x)$ is not a coherent state,
  some $f_k$ with $k>0$ does not vanish and we have $x_{\rm max}>0$.
  Also for the rotated Bargmann function,
  ${\mathcal R}_{x_{\rm max}}\circ \psi(x)$,
  there exist some nonvanishing coefficients,
  $\tilde{f}_{k} [x_{\rm max}]$, since
  rotating a non-coherent
  state results still in a non-coherent state. At least
  one of the non-vanishing coefficients must be negative
  $f_{k} [x_{\rm max}]<0$. If no coefficients were
  negative,
  there would be a  rotation ${\mathcal R}_x$ that increases $f_0[x]$
  further.

  Now define
  \begin{equation}
    |\psi'\rangle={\mathcal F}_2\circ{\mathcal R}_{x_{\rm max}}(|\psi\rangle).
  \end{equation}
  We have then have
  $W^{(q)}_{|\psi'\rangle} \ge W^{(q)}_{|\psi\rangle}$.
  Only if  the
  signs of coefficients  of ${\mathcal R}_{x_{\rm max}}\circ\psi(\gamma)$
  are related according to
  $\tilde{f}_k[x_{\rm max}] = |\tilde{f}_k[x_{\rm max}]|(-1)^k$
  we have $W^{(q)}_{|\psi'\rangle} =W^{(q)}_{|\psi\rangle}$
  according to lemma \ref{lemma1}.
  In this case one may replace the
  rotation ${\mathcal R}_{x_{\rm max}}$ by
  ${\mathcal R}_{x_{\rm max}-\epsilon}$
  for sufficiently small $\epsilon$.
  For small $\epsilon>0$ we have
  $\tilde{f}_1[x_{\rm max}-\epsilon]>0$ since 
  $\tilde{f}_0[x]$ has an isolated maximum
  at $x=x_{\rm max}$ and
  $\frac{d}{dx}\tilde{f}_0[x]=\tilde{f}_1[x]$. We can chose
  $\epsilon$ small enough such that the
  negative coefficients remain negative (we have shown above
  that at least one coefficient is negative).
  Since now $\tilde{f}_0[x_{\rm max}-\epsilon]>0$ and 
  $\tilde{f}_1[x_{\rm max}-\epsilon]>0$
  the relation
  $\tilde{f}_k[x_{\rm max}-\epsilon] 
  \neq|\tilde{f}_k[x_{\rm max}-\epsilon]|(-1)^k$
  cannot be fullfilled and we have
  $W^{(q)}_{|\psi'\rangle} >W^{(q)}_{|\psi\rangle}$.
\end{proof}

This proof shows what a possible
future proof of Lieb's conjecture on $S_{|\psi\rangle}$ might be.
With
$S_{|\psi\rangle}=2j-\sum_{n=2}^\infty \frac{Y^{(n)}_{|\psi\rangle}}{n(n+1)}$
Lieb's conjecture is equivalent to the statement that all dual moments
$Y^{(n)}$ are maximalized by coherent states for $n=2,3,\dots$. For
example, if
it can be shown that $Y^{(n)}_{{\mathcal F}_2(|\psi'\rangle)}
\ge Y^{(n)}_{|\psi\rangle}$ for $n=2,3,\dots$ an analogous proof
can be given for Lieb's conjecture.

It might be also interesting
to consider the delocalized pure states characterized by the {\sl
maximal entropy}.
In the space of mixed quantum states,
the maximally delocalized state is proportional to the identity matrix,
$\rho_*:={\mathbb I}/(2j+1)$, for which
$S^{(q)}_{\rho_*}=\log (2j+1)$ and $R_{\rho_*}= 2j+1$.
However, the maximal entropy $S^{\rm max}$ is in general unknown, if one
looks for a maximum in the space of all pure states.
Lee conjectured \cite{lee:88} that this maximum is achieved for pure
states with possibly regular
distribution of all $2j$ zeros of the Husimi function on the sphere.
Such a distribution of zeros is easy to specify for
$2j=4,6,8,12,20$, which correspond to the Platonic polyhedra
\cite{zyczkowski:2001},
and it seems to be plausible that these states provide the maxima for
all the generalized entropies $S^{(q)}$.

\subsection{Generalizations to other compact phase spaces}
\label{subsec:generalizations}

We have introduced generalized Lieb--Wehrl
entropies
and discussed their minimal values (Lieb--Wehrl conjecture).
The above results, obtained for the
Bloch sphere $S^2$ and the $SU(2)$ spin coherent states,
may be generalized for classical phase spaces
associated with other (compact) Lie groups such as
$SU(d)$; $d\ge 2$.
In that case the classical phase spaces are
equivalent to complex projective spaces,
 ${\mathbb CP}^{d-1}$, and
the dimension of the Hilbert space equals
\footnote{This is true for the physically 
important class of irreducible representations of the group $SU(d)$ 
on Hilbert spaces $V$ \cite{gnutzmann:98}.}
$N=\dim V=\binom{m+d-1}{m}$,
with $m=1,2,\dots$.
The Wehrl entropy of an $SU(d)$
coherent state equals \cite{slomczynski:98,jones:90}
\begin{equation}
  S_{\rm coh}= m \left[ \Psi \left( m+d\right)
    -\Psi
    \left( m+1\right)
  \right],
  \label{meandd}
\end{equation}
which for $d=2$ reduces to $(N-1)/N$ as desired.
Our generalization of the Lieb--Wehrl conjecture
stated that this value gives the
minimal Wehrl entropy for all pure states
corresponding to the classical phase space ${\mathbb CP}^{d-1}$.
The methods we used to give proofs of
some special cases of this conjecture on the sphere should
be applicable in that case as well.

\section{Random pure states}
\label{sec:random}

\subsection{R{\'e}nyi-Wehrl entropies for random states}
\label{subsec:randRW}

Spin coherent states are localized in phase space as much as
allowed by the uncertainty principle and are characterized by
the minimal R{\'e}nyi--Wehrl entropies.
Let us emphasize that for $j \gg 1$ the pure states exhibiting
Wehrl entropy of the order of $S_{\rm min}$ are not typical.
In a typical situation the density of the zeros of the
Bargmann or Husimi function is close to uniform on the sphere
\cite{leboeuf:90},
and the Wehrl entropy of such delocalized pure states is
found to be large.

In order to analyze,  to what extend the numbers
obtained in (\ref{eq:coherententropies}) are small, we compute
the mean entropies averaged over the ensemble of random pure states.
A random pure state $|\Phi\rangle$, can be generated according to the
natural uniform measure on the space of pure
states by taking $|\Phi\rangle =U|\kappa\rangle$, where
$|\kappa\rangle=|j\rangle$
is the reference state and $U\in U(N)$ with $N=2j+1$ is a
random unitary matrix distributed according to the Haar measure
$d\mu(U)$. Such random matrices
pertain to the {\sl circular unitary ensemble} (CUE) \cite{mehta:91},
often used to describe quantum chaotic systems \cite{haake:2000}.
We compute the mean R{\'e}nyi--Wehrl entropies of $N$-dimensional pure
states, $\langle S^{(q)}q\rangle$ by taking the average
with respect to this measure. Let us start computing the mean values
of the $q$-th moments
\begin{equation}
  \left\langle W^{(q)} \right\rangle :=
  \,\int_{U(N)}\,\left(
    N \int_{S^2}\!\!  [{\mathcal H}_{|\Phi\rangle} (\gamma)]^q
    d\mu(\gamma)
  \right)\, d\mu (U).
\label{rand1}
\end{equation}
Since
${\mathcal H}_{|\Phi\rangle} (\gamma)=|\langle \Phi |\gamma \rangle|^2 =
|\langle j |U^{-1} {\mathcal R}(\gamma,0)  |j\rangle|^2$
one may
interchange the order of integration and use the
invariance of the Haar measure \cite{slomczynski:98}.
Putting $V:=U^{-1}{\mathcal R}(\gamma,0)$ we conclude that
\begin{equation}
  \left\langle W^{(q)} \right\rangle=
  N \int_{U(N)} |\langle j |V|j
  \rangle |^{2q}  d\mu (V),
  \label{rand2}
\end{equation}
since $\int d\mu(\gamma)=1$.

This average may be computed based on earlier results of
Ku\'{s} et al. \cite{kus:88} or Jones \cite{jones:90}. We obtain
$\langle W^{(q)}\rangle = N\Gamma(N)\Gamma(q+1)/\Gamma(q+N)$,
which substituted into the definition leads to
\begin{equation}
  \left\langle S^{(q)} \right\rangle_N =
  {\frac{1}{1-q}} \ln \bigl[ \Gamma(N+1) \frac{\Gamma(q+1)}{\Gamma(q+N)}
  \bigr].
  \label{RWrand}
\end{equation}
Thus the mean Wehrl participation number of a typical random state is
\begin{equation}
  R^{\rm rand}=e^{S_2}=\frac{N+1}{2}.
\label{Rrand}
\end{equation}
In the limit $q \to 1$ formula (\ref{RWrand}) gives the
mean Wehrl entropy of a random pure state
\cite{slomczynski:98}
\begin{equation}
  \left\langle S\right\rangle_N=\Psi \left( N+1\right) -\Psi \left(
    2\right) =\sum_{n=2}^{N}{\frac{1}{n}},  \label{wehmean}
\end{equation}
where $\Psi(x)=\Gamma '(x)/\Gamma(x)$ denotes the digamma function.
Thus the mean Wehrl entropy of a random state is equal to the
Shannon entropy \cite{jones:90} of its expansion
in a relatively random basis \cite{jones:90,kus:91}.
The mean Wehrl entropy
was used to obtain bounds for dynamical entropies characterizing the
properties of quantum maps \cite{slomczynski:98,mirbach:95,mirbach:98}.
Note that another normalization
of the coherent states used in  \cite{slomczynski:98},
leads to results shifted by
a constant $- \ln N$. These normalization
allows one for a direct comparison between the entropies
describing the states of various $N$.

In the asymptotic limit $N\rightarrow \infty $ the mean
entropy $\langle S \rangle_N$
behaves  as $\ln N+\boldsymbol{\gamma}
-1\approx \log N-0.42278$, which is close
to the maximal possible Wehrl entropy
for mixed states $S_{\rho _{\ast }}=\log N$.
Here $\boldsymbol{\gamma}$ denotes the Euler constant.

Let us mention that the above technique of computing the averages over
the CUE does not work for the unitary symmetric matrices of circular
orthogonal ensemble, (COE),  since writing
$V:=(UU^T)^{-1}{\mathcal R}_{\gamma}$
one cannot perform the average in  (\ref{rand2}).
The mean Wehrl entropy of a random 'symmetric' state,
$|\Phi_s\rangle:=U^TU|j\rangle$ differs therefore
from the Shannon entropy $S_o$ of an eigenvector of a symmetric random
unitary matrix typical of COE,
$S_o=\Psi ( N/2+1) -\Psi (3/2)\sim$ $\ln N +\gamma+\ln2-2
\approx \ln N -0.7296$
\cite{jones:90}. Numerical computations show
that the mean Wehrl entropy of random vectors $|\Phi_s\rangle $ is
slightly smaller than the entropy of
$|\Phi\rangle$ given by (\ref{wehmean}),
and this difference vanishes in the asymptotic limit $N \to\infty$
\cite{zyczkowski:2001}.

\subsection{Distance between Husimi functions and uniform
distribution}
\label{subsec:distance}

The theory of {\sl quantum ergodicity}
deals with the semiclassical
properties of eigenfunctions
of a Laplace operator. The Shnirleman theorem says
that the expectation values of quantum observables
in eigenstates
tend to the mean values of the corresponding classical quantities
if the classical system is ergodic \cite{shnirelman:74}.
In the semiclassical limit almost all eigenfunctions
of quantum chaotic billiards tend, in a weak sense,
to the measures covering uniformly entire
configuration space available (the domain of the billiard) \cite{zelditch:96}.
An analogous statement concerning the phase space
implies that for almost all eigenstates the corresponding Wigner
(or Husimi) distributions condensate uniformly on the energy surface.

We may thus expect that the Husimi distribution of a random
eigenstate will tend to the uniform distribution,
$(2j+1){\mathcal H}_*(\gamma)=1$ in the
limit $j\to \infty$\footnote{In this limit
$p(\gamma)=(2j+1){\mathcal H}(\gamma)$ may
be interpreted as a classical probability distribution on the
sphere.}.
Since strong
convergence is excluded by the presence of the $2j$ zeros of the Husimi
distribution for any pure state, we
will consider weak convergence only.
To characterize this convergence quantitatively we introduce the
$L_2$ distance
\begin{equation}
  L_2({\mathcal H}_{|\Phi\rangle}(\gamma),{\mathcal H}_{*}(\gamma)):=
  \Bigl( (2j+1) \int_{S^2}
  [{\mathcal H}_{|\Phi\rangle}(\gamma) - {\mathcal H}_{*}(\gamma)]^2
  d\mu(\gamma)
  \Bigr)^{1/2}.
  \label{distd2}
\end{equation}
We may express the distance to the
uniform distribution by the Wehrl participation number \eqref{eq:PRdef}
\begin{equation}
  L_2^2({\mathcal H}_{|\Phi\rangle},{\mathcal H}_{*})=
  \frac{2j+1}{4 \pi} \int_0^{2\pi}d\varphi
  \int_0^{\pi}\sin\vartheta d \vartheta
  \Bigl[ |\langle \gamma|\Phi\rangle|^4 - \frac{1}{(2j+1)^2}
  \Bigr]
  = \frac{1}{R} -\frac{1}{2j+1}.
  \label{distd22}
\end{equation}
Using the previous result (\ref{Rrand}) we get
\begin{equation}
  \langle
  L_2^2({\mathcal H}_{|\Phi\rangle},{\mathcal H}_{*}) \rangle =
  \frac{2j}{ (2j+1)(2j+2)} \to  0
  \qquad {\rm as} \qquad j\to \infty ,
  \label{distran}
\end{equation}
where the average is taken over the natural, unitarily invariant,
measure on the manifold of pure states.

In other words, in the semiclassical limit ($j \to \infty$)
the Husimi distribution of a typical random state
tends  to the uniform distribution of the sphere in the weak sense.
On the other hand, applying (\ref{eq:PRdef}) we see that
 the analogous distance
for the localized coherent state tends to a constant
\begin{equation}
L_2^2({\mathcal H}_{\rm coh},{\mathcal H}_{*})= \frac{2j}{4j+2}
  \to \frac{1}{2},
  \qquad {\rm as} \qquad j\to \infty ,
\label{distcoh}
\end{equation}
which emphasizes the intuitive fact that the coherent states
are exceedingly non-typical in the space of pure states of a large
dimensional Hilbert space.

\section{Concluding remarks}

The Wehrl entropy of a pure quantum state may be considered as a useful
measure of its localization in classical phase space.
In contrast with the
so-called 'eigenvector statistics', it is defined without any ambiguity,
provided the quantum problem admits to introduce the set of
coherent states distinguished
by the corresponding classical dynamics. In the analogy to the
well known notion of the {\sl R{\'e}nyi entropies} \cite{Re61},
we defined and investigated the
R{\'e}nyi--Wehrl entropies, which may serve as complementary
measures of localization. Although the family of generalized entropies
is parametrized by the continuous real R{\'e}nyi parameter $q$,
the generalized entropies are most easily evaluated for integer
values of $q$.

Every pure state of a $N$--dimensional Hilbert space may be
uniquely represented by the set of $N-1$ zeros of its Husimi
 (Bargmann)
functions  defined by the family of coherent states. For the
physically most relevant case of the $SU(2)$ coherent states,
any state can be represent by $N=2j$ indistinguishable points on the
sphere (which may coalesce) \cite{majorana:32,bacry:74,penrose:89}.
In this {\sl stellar representation} the coherent states are
distinguished by being the only states, for which all
zeros are located in a single point.
It is thus natural to expect, that coherent states
mimimize the Wehrl entropy,
as conjectured by Lieb \cite{lieb:78}.

Not being able to prove this conjecture in its full glory,
in this work we:
\begin{itemize}
\item[i]) have shown that coherent states provide local minima
  (maxima)
  of the generalized entropies (moments) for any value
  of the R{\'e}nyi parameter $q >0$;
\item[ii]) have proved the generalized Lieb conjecture for integer values
  $q=2,3,...$.
\end{itemize}
To achieve these  goals we introduced
and used the notion of (non-unitary) entropy-reducing maps
in the space of pure states, which might be helpful
in further attempts to prove the Lieb conjecture.
Moreover, in this paper we analyzed the random pure states
and computed the values of generalized Wehrl entropies
averaged over the natural, unitarily invariant measure in the
manifold of pure states of $N$--dimensional Hilbert space.

We are indebted to Wojciech S{\l}omczy{\'n}ski
for numerous suggestions, long lasting interaction
related to this project and reading of the manuscript.
We have also enjoyed discussions with Christoffer Manderfeld and
  Marek Ku\'s.
  This work has been supported by the Minerva Foundation
and the Polish grant of Komitet Bada{\'n} Naukowych
number  2P03B 072 19.

\appendix

\section{Rotations of Bargmann functions}
\label{app:rotations}

Rotations of a state $|\psi\rangle$ by some ${\mathcal R}\in SU(2)$ corresponds
to transformations of the corresponding Bargmann and Husimi function.
In the defining representation of $SU(2)$ by $2\times 2$ matrices ($j=1/2$)
almost any rotation ${\mathcal R}\in SU(2)$ (with respect to Haar measure)
can be written as in the form
\begin{equation}
  {\mathcal R}_{1/2}(\alpha,\phi)=
  \begin{pmatrix}
    \frac{e^{i\frac{\phi}{2}}}{\sqrt{1+|\alpha|^2}} &
    -\frac{\overline{\alpha}\,e^{-i\frac{\phi}{2}}}{\sqrt{1+|\alpha|^2}} \\
    \frac{\alpha\, e^{i\frac{\phi}{2}}}{\sqrt{1+|\alpha|^2}} &
    \frac{e^{-i\frac{\phi}{2}}}{\sqrt{1+|\alpha|^2}}
  \end{pmatrix}
  \label{eq:rotmat}
\end{equation}
for some complex $\alpha$ and angle $\phi$.
This looks much nicer using the Gaussian decomposition
\begin{equation}
  {\mathcal R}(\alpha,\phi)=
  \begin{pmatrix}
    1 & 0 \\
    \alpha & 1
  \end{pmatrix}
  \begin{pmatrix}
    (1+|\alpha|^2)^{-1/2}e^{i\frac{\phi}{2}} & 0\\
    0 & (1+|\alpha|^2)^{1/2} e^{-i\frac{\phi}{2}}
  \end{pmatrix}
  \begin{pmatrix}
    1 &-\overline{\alpha}e^{-i \phi}\\
    0 & 1
  \end{pmatrix}       \ .
\end{equation}
In the general case of the $2j+1$ dimensional Hilbert space $V_j$
the rotation ${\mathcal R}(\alpha,\phi)$ is represented by the
operator
\begin{equation}
  {\mathcal R}_{j}(\alpha,\phi)=
  e^{\alpha J_-} e^{(i\phi-\log(1+|\alpha|^2))J_z}
  e^{-\overline{\alpha}J_+}   \ .
  \label{eq:representrot}
\end{equation}
We omitted the index $j$ of this operator throughout the text.

Rotating a coherent state gives another coherent state (up to a
normalization constant and a phase factor).
\begin{align}
  {\mathcal R}(\alpha,\phi)|\gamma)=&
  \exp{
    \left( \left({\mathcal R}(\alpha,\phi)\circ\gamma\right) J_-\right)
    }\times
  \nonumber\\
  &\times\exp{
    \left(
      \left[i\phi+{\rm log}\left(
          \frac{(1-\overline{\alpha}\gamma e^{-i\phi})^2}{1+|\alpha|^2}\right)
      \right] J_z
    \right)}|j\rangle_j\\
  =&\frac{(e^{i \phi}-\overline{\alpha}\gamma)^{2j}}{(1+|\alpha|^2)^j}
  e^{-i\phi j}
  |{\mathcal R}(\alpha,\phi)\circ\gamma),
\end{align}
where
\begin{equation}
  {\mathcal R}(\alpha,\phi)\circ\gamma=
  \frac{\gamma+\alpha e^{i\phi}}{e^{i\phi}-\overline{\alpha}\gamma}.
  \label{eq:rotgamma}
\end{equation}
If a rotation acts on a state $|\psi\rangle \mapsto {\mathcal R}(\alpha,\phi)
|\psi\rangle$ its Bargmann function transforms like
\begin{equation}
  \psi(\gamma)=\langle \psi |\gamma) \mapsto \psi'(\gamma)
  =\langle {\mathcal R}(\alpha,\phi) \psi|\gamma)=
  \langle\psi|{\mathcal R}(\alpha,\phi)^\dagger|
  \gamma)
\end{equation}
where
\begin{equation}
  {\mathcal R}(\alpha,\phi)^{\dagger}={\mathcal R}(-\alpha e^{i\phi},-\phi).
\end{equation}
The rotated Bargmann function thus is
\begin{equation}
  \psi'(\gamma)=\langle\psi|{\mathcal R}(-\alpha e^{i\phi},-\phi)|\gamma)=
  \frac{(1+\overline{\alpha}\gamma)^{2j}}{(1+|\alpha|^2)^j}e^{-i\phi j}
  \psi({\mathcal R}(-\alpha e^{i\phi},-\phi)\circ\gamma) \ .
\end{equation}
In the stellar decomposition it is obvious that the
rotation of the state corresponds to rotation of the
zeroes of its Bargmann function
\begin{align}
  \psi(\gamma)={\mathcal N}^{-1}\prod_{k=1}^{2j}
  \left(\gamma-\gamma^{(0)}_k
  \right)\mapsto
  \psi'(\gamma)={{\mathcal N}'}^{-1}\prod_{k=1}^{2j}
  \left(\gamma- \mathcal{R}(\alpha,\phi)\circ \gamma^{(0)}_k\right) ,
\end{align}
where
\begin{equation}
  {\mathcal{N}}'={\mathcal N}(1+|\alpha|^2)^j e^{-i\phi j} \prod_{k=1}^{2j}
  \left(1-\gamma^{(0)}_k\overline{\alpha}e^{-i\phi}\right)^{-1}.
\end{equation}
These transformations are well behaved
if some zeros are on the south-pole ${|\gamma^{(0)}_k|~\rightarrow~\infty}$.

\bibliographystyle{amsplain}

\end{document}